\documentclass[conference]{IEEEtran}

\usepackage{newtxtext}            % Times-like text
\usepackage{newtxmath}            % Times-like math + AMS symbols
\usepackage[scaled=.90]{helvet}   % Helvetica for \sffamily
\usepackage{microtype}            % Better spacing, no bitmap fonts
\IEEEoverridecommandlockouts
\usepackage{cite}

\usepackage{amsmath}

\usepackage{algorithmic}
\usepackage{graphicx}
\usepackage{textcomp}
\usepackage{xcolor}
\usepackage{url}

\def\BibTeX{{\rm B\kern-.05em{\sc i\kern-.025em b}\kern-.08em
    T\kern-.1667em\lower.7ex\hbox{E}\kern-.125emX}}
\begin{document}

\title{HyProv: Hybrid Provenance Management for Scientific Workflows

}

\author{
\IEEEauthorblockN{Vasilis Bountris}
\IEEEauthorblockA{\textit{Department of Computer Science} \\
\textit{Humboldt-Universität zu Berlin} \\
Berlin, Germany} 
\and
\IEEEauthorblockN{Lauritz Thamsen}
\IEEEauthorblockA{\textit{School of Computing Science} \\
\textit{University of Glasgow} \\
Glasgow, United Kingdom}
\and
\IEEEauthorblockN{Ulf Leser}
\IEEEauthorblockA{\textit{Department of Computer Science} \\
\textit{Humboldt-Universität zu Berlin} \\
Berlin, Germany}
}

\maketitle

\begin{abstract}
Provenance plays a crucial role in scientific workflow execution, for instance by providing data for failure analysis, real-time monitoring, or statistics on resource utilization for right-sizing allocations. The workflows themselves, however, become increasingly complex in terms of involved components. Furthermore, they are executed on distributed cluster infrastructures, which makes the real-time collection, integration, and analysis of provenance data challenging. Existing provenance systems struggle to balance scalability, real-time processing, online provenance analytics, and integration across different components and compute resources. Moreover, most provenance solutions are not workflow-aware; by focusing on arbitrary workloads, they miss opportunities for workflow systems where optimization and analysis can exploit the availability of a workflow specification that dictates, to some degree, task execution orders and provides abstractions for physical tasks at a logical level.

In this paper, we present HyProv, a hybrid provenance management system that combines centralized and federated paradigms to offer scalable, online, and workflow-aware queries over workflow provenance traces. HyProv uses a centralized component for efficient management of the small and stable workflow-specification-specific provenance, and complements this with federated querying over different scalable monitoring and provenance databases for the large-scale execution logs. This enables low-latency access to current execution data. Furthermore, the design supports complex provenance queries, which we exemplify for the workflow system Airflow in combination with the resource manager Kubernetes. Our experiments indicate that HyProv scales to large workflows, answers provenance queries with sub-second latencies, and adds only modest CPU and memory overhead to the cluster.

\end{abstract}

\begin{IEEEkeywords}
  Provenance, Workflow Engines, Cluster Resource Management, Online Query Processing, Scalability
\end{IEEEkeywords}

\section{Introduction}
\label{sec:intro}

Scientific Workflows have become a popular tool across scientific domains for addressing complex and large-scale data analysis problems~\cite{yatesReproduciblePortableEfficient2021, lehmannFORCENextflowScalable, schaarschmidtWorkflowEngineeringMaterials2022}.
As workflows grow in complexity and data volumes increase—both in terms of input and generated output—they provide techniques for enhancing reproducibility, scalability, and fault tolerance \cite{liewScientificWorkflowsMoving2016}.
These capabilities are enabled and managed by Scientific Workflow Management Systems (SWMSs), which handle tasks such as scheduling, managing dependencies between tasks, interfacing with the underlying infrastructure for execution, restarting failed tasks, and collecting and presenting provenance of the workflow engine. \cite{perezSystematicReviewProvenance2018a}.

As workflows demand increasing computing and storage resources, SWMSs simplify the use of clusters by integrating with resource managers (RMs) such as Kubernetes\footnote{\url{https://kubernetes.io/}} or Slurm\footnote{\url{https://slurm.schedmd.com/}}. In doing so, workflow engines deliberately abstract away execution details to improve portability and usability for developers. Such a design implies that each workflow execution interacts with multiple underlying components, including RMs, operating systems, container managers, distributed storage systems, file systems, and task runtimes. Each of these components generate provenance at different locations, volumes, formats, and frequencies, making it difficult to obtain a unified and consistent picture of workflow executions.

Provenance in the context of workflow systems encompasses a wide variety of concepts, ranging from a particular SW's task history to the performance characteristics of a workflow run, the utilization of cluster resources and the workflow versioning itself~\cite{khanSharingInteroperableWorkflow2019}. Effective provenance capture and analysis can support the development, debugging, testing, and execution of a SW~\cite{davidsonProvenanceScientificWorkflows}.
Provenance data is usually multimodal, encompassing various types such as time-series data, logs, execution traces, and configuration files.
Some components also generate large volumes of data, such as detailed task execution logs and high-frequency time-series performance metrics, while others produce only small amounts, like schedulers and workflow execution engines.
This variability presents significant challenges for scalable and flexible provenance management.

An effective provenance management system should enable researchers to perform \emph{complex, workflow-aware queries} across the combined provenance data of all components, providing a unified and transparent view of workflow execution.
By complex we mean queries that span the abstraction layers of the workflow and the infrastructure and are not trivially derivable by existing workflow managers or monitoring systems of the cluster.
This capability must be available not only after the workflow has been completed but also \emph{online} (during execution), facilitating immediate insights, debugging information, and adjustments.
As SW execution continues to expand in scale and complexity, the system should accommodate this growth without sacrificing \emph{scalability} or introducing significant \emph{overhead}.

To address these requirements, there are two primary approaches to provenance data management: centralized storage or federated queries.
Most systems aggregate provenance data into a central data store, such as a provenance warehouse, simplifying querying but introducing significant maintenance overhead, data duplication, and scalability issues during the capture process as workflows grow in size and complexity~\cite{herschelSurveyProvenanceWhat2017b}. Furthermore, centralized systems struggle to always have the most up-to-date information available, which hinders their usage for online monitoring or debugging.
In contrast, a federated approach keeps provenance data distributed across the nodes and systems of the cluster, integrating data only when necessary to answer specific user requests.
When properly designed, this method reduces maintenance, is easier to extend, and scales more effectively.
However, it introduces challenges in online data integration and efficient query execution, as integrating data from diverse sources on demand is complex.
Presumably due to such concerns, federated provenance management has not received much attention so far. 

Another restriction of existing provenance systems is that they are typically general-purpose, i.e., designed for arbitrary workloads and thus workload-agnostic~\cite{perezSystematicReviewProvenance2018a}. Accordingly, they miss opportunities for query optimization and user support when applied to workflow systems, where the relationships of prominent execution objects to query, such as tasks, workflows, and dependencies, are actually predefined by a workflow specification. Workflow-aware provenance systems should be able to exploit this information for easier and faster query execution.

In this paper, we propose HyProv, a new hybrid system for managing heterogeneous provenance data that combines the strengths of the central and federated approaches to achieve online, scalable, and workflow-aware provenance queries. Our design is based on two key decisions. First, we focus on integrating the workflow management system provenance and the infrastructure provenance. This integration reconnects the abstracted workflow information with the physical execution objects by tracking the events emitted by these components. Second, we propose a hybrid architecture consisting of a central store for workflow-specific information and a federated component that leverages the established ecosystem of monitoring databases and tools. This allows steering query execution efficiently from the central store while removing the need to create new online functionality or scalable infrastructure from scratch (see Figure~\ref{fig:system_overview}). To this end, HyProv introduces a lightweight intermediate layer that translates workflow-related queries into a mixture of centralized specification queries and decentralized infrastructure-level queries. This layer allows scientists and workflow developers to express complex queries in familiar terms, while the system handles the retrieval and aggregation of provenance data from various distributed sources. Specifically, our work makes the following contributions:
\begin{itemize}
    \item We present a novel system design that integrates centralized workflow and decentralized infrastructure provenance data in a hybrid manner.
    \item We implement HyProv for the workflow system Airflow and the distributed systems Kubernetes, Prometheus, and ElasticSearch.
    \item We evaluate HyProv's workflow-awareness, online querying, scalability, and overhead experimentally using our prototype implementation, a commodity cluster, and synthetic workflows.
\end{itemize}

The paper is organized as follows: Section~\ref{sec:related} reviews related work, and Section~\ref{sec:background} summarizes key concepts in provenance management.
Section~\ref{sec:system} details our system's design, components, data model, and operation.
% , supported by diagrams for clarity.
Section~\ref{sec:prototype} presents the implemented prototype.
Section~\ref{sec:experiments} presents selected experiments that showcase the effectiveness of our approach.
Finally, Section~\ref{sec:conclusion} concludes the paper with a summary of our findings and discusses potential directions for future research.

\section{Related Work}
\label{sec:related}
Work on provenance management followed a range of approaches, from stand-alone provenance stores to systems integrated within SWMSs. Several solutions have been specifically developed for High-Performance Computing (HPC).

One typical approach, exemplified in an end-to-end manner for the HPC Workflow System Pegasus~\cite{deelmanPegasusWorkflowManagement2015,papadimitriouEndendOnlinePerformance2021}, is to design an architecture that aggregates performance statistics from various sources, correlates provenance between different layers, and publishes them all in a centralized store.
A part of the architecture above, Stampede~\cite{vahiGeneralApproachRealTime2012}, has been adapted for the WFMS Triana~\cite{harrisonWSRFWorkflowTriana2008}, employs a 3-layered model (workflow-infrastructure model, log aggregators and loaders, querying interface) to store and make available the provenance as normalized logs in an SQL archive.
MIDA \cite{souzaLightweightDataIntegration2023a} uses an adapter-based approach for monitoring multiple workflows concurrently and stores everything in a DBMS. ProvLake~\cite{souzaEfficientRuntimeCapture2019a} similarly integrates provenance from various sources and stores it into a multi-modal knowledge graph.
Komadu~\cite{suriarachchiKomaduCaptureVisualization2015} follows a similar approach but focuses on data provenance.
All these systems have low overhead but do not address adequately the scalability issues that can arise with complex provenance queries. 

Some WFMS have built-in provenance capabilities.
Nextflow~\cite{ditommasoNextflowEnablesReproducible2017}, a widely used WFMS in genomics, focus on reproducibility with post-execution summaries of resource usage. 
However, it lacks real-time, queryable interfaces and detailed tracking of dependencies.
Similarly, Apache Airflow\footnote{\url{https://airflow.apache.org/}}, a Python-based WFMS, uses SQL to allow tracking task states and dependencies but does not support complex, workflow-aware provenance queries.

Provenance management solutions that employ federated approaches have seen limited application. SPADE~\cite{gehaniEfficientQueryingDistributed2010} is such an approach, focused solely on data provenance, that relies on a decentralized system with each node maintaining a repository of the provenance gathered on it. While not an implemented provenance management system, in \cite{ellqvistUsingMediationAchieve2009} the authors explore building a global provenance schema and query mediation to integrate and query provenance in three different subsystems.

Existing systems lack support for online, workflow-aware provenance querying in a scalable manner. HyProv addresses this gap by integrating workflow and infrastructure provenance within a hybrid architecture.

\section{Provenance in Scientific Workflow Executions}
\label{sec:background}
This section provides an overview of the key aspects of provenance management for SWs, including an introduction to workflow systems, the generation of provenance data, and the current approaches for querying and analyzing this information.

\subsection{Scientific Workflows }
SWs serve as structured representations for orchestrating data analysis and computational experiments in scientific research~\cite{deelmanWorkflowsEScienceOverview2009}. They are typically modeled as Directed Acyclic Graphs (DAGs), where nodes represent tasks and edges denote dependencies. This structured abstraction facilitates the organization of complex data-processing pipelines, ensuring that computations proceed according to predefined logical or data-driven relationships. 

\subsection{Scientific Workflow management Systems}

Scientific Workflow Management Systems (SWMSs) realize these abstractions in practice. They streamline the development, execution, and monitoring of workflows across various infrastructures, from high-performance computing clusters to cloud-based platforms. SWMSs are designed to handle the growing computational demands of modern SWs, including data-intensive applications requiring vast processing and storage resources. By abstracting away much of the complexity associated with distributed computing, SWMSs allow scientists to focus on their domain-specific challenges rather than infrastructure management. They typically offer a certain level of portability across different systems, user-friendly interfaces (graphical or command-line), and features such as parallel execution of tasks, dynamic resource allocation, and provenance capture~\cite{liuSurveyDataIntensiveScientific2015a}. Popular frameworks include Pegasus and Apache Airflow, which provide robust environments to manage and execute workflows effectively.

\subsection{Workflow Execution}
For executing a workflow, SWMS work together with other infrastructure components, such as RMs or file systems. Workflow execution is the process of translating a workflow’s task definitions and dependencies into a series of executable steps within the computational infrastructure. 
The execution process begins with the SWMS constructing an initial DAG that represents all executable tasks and their dependencies.
Tasks at the start of the DAG, without any prerequisites, are ready-to-run immediately, while other tasks have to wait until all required preceding tasks are completed.
The SWMS interfaces with an RM — such as Kubernetes or Slurm — to communicate resource requirements, requesting and assigning CPU, memory, storage, and other necessary resources for each task.
The SWMS also keeps track of tasks' state while they are taken over by the infrastructure, and may decide to retry in cases of failure.
The RM is of central importance to a workflow execution, but numerous more components are involved in a workflow's execution, like a (distributed) file system that handles workflow data and fault tolerance, container runtimes, and monitoring databases that keep track of the state of the cluster and the applications that run inside the tasks.

Another critical aspect of workflow execution, especially in the context of provenance, is the automatic parallelization feature provided by some SWMSs. Systems like Nextflow allow the definition of abstract tasks, which represent high-level units of computation. These tasks are associated with groups of input files or parameters, enabling the SWMS to dynamically map these inputs to concrete tasks during runtime. This means that a single abstract task can generate an a priori unknown number of concrete tasks, each executing independently on different input files or data subsets.
This approach offers scalable execution with minimal user effort, as tasks can be distributed across multiple compute resources for parallel processing. However, it introduces additional complexity in mapping and tracking the defined workflow to its runtime representations. 

\subsection{Provenance Generation}
Provenance data in workflow environments is produced by many components, each recording events at different levels of detail, in varying formats, and at very different volumes. Workflow management systems typically log high-level execution information such as task dependencies, parameters, and input/output files. In contrast, resource managers (e.g., Kubernetes, Slurm) and infrastructure services capture low-level details such as task-to-node assignments, job states, and failure reports. Since each component produces provenance independently, the resulting information is highly heterogeneous and often siloed.

Similarly, the other interacting components within the infrastructure also produce valuable workflow provenance data that can support understanding a workflow's execution:

\begin{itemize}
    \item \textbf{Execution Objects}: Different infrastructure environments represent tasks through specific execution objects. For instance, in a Kubernetes-managed workflow, tasks are executed within \textit{pods}, while in Slurm, tasks are run as \textit{jobs} on designated nodes. Capturing these execution objects and their attributes, such as physical location or assigned/used resources, provides insight into how tasks are deployed and managed in different environments.
    
    \item \textbf{Process Details}: On each node, tasks may spawn multiple processes to perform computations. Recording identifiers and state changes of these processes (e.g., process IDs, start/stop times, exit codes) provides fine-grained provenance useful for debugging and reproducibility. 
    
    \item \textbf{File Access, Data Usage, and Transfers in Distributed File Systems}: Provenance information may include details about files or datasets a task is accessing, such as file paths, data volume, and timestamps. In workflows utilizing distributed file systems like CEPH or Lustre, provenance data also encompasses records of data transfers between nodes, storage locations, and access details. This involves tracking which nodes store specific data blocks, which tasks access or transfer these blocks, and the protocols employed for data replication and redundancy management. By including information on data sharding, replication factors, and file caching, provenance data ensures robust tracking of data lineage and supports reliable monitoring of data movement, which is crucial for understanding workflows requiring intensive data sharing and processing across multiple nodes.

    \item \textbf{Error and Alert Logs}: Infrastructure components, such as the node's operating systems and RMs, generate logs indicating any errors or alerts during task execution. Including these logs in provenance data allows for tracking failures and understanding their context within the workflow execution.
    
    \item \textbf{Resource Utilization Metrics}: Provenance data often includes detailed records of resource consumption, such as CPU and memory usage, throughout a task's execution. These metrics, typically captured by monitoring tools or the RM, are invaluable for understanding a task’s computational demands and for optimizing future workflow runs. For instance, tracking CPU spikes or high memory usage can highlight resource bottlenecks or inefficiencies, enabling better resource allocation in subsequent executions.
\end{itemize}

Each of these types of provenance information provides essential insights but is typically siloed within individual components of the infrastructure. How this data is stored has a major impact on its later usability: while many systems keep provenance locally in logs or monitoring databases, effective provenance management requires bringing these heterogeneous records into a form that supports integrated, workflow-aware querying.

\subsection{Provenance Queries}
Provenance queries enable users to retrieve detailed information about the origins, dependencies, transformations, and resource usage of tasks and entire workflows. They allow both real-time monitoring during execution and retrospective analysis across past workflow runs.

Such queries can be broadly divided into \emph{online} and \emph{offline}. Online queries operate while a workflow is actively running, providing up-to-date information for debugging and performance monitoring. They are typically limited in scope to a single workflow and must be served with low latency, since their value lies in enabling quick reactions to emerging issues. Offline queries, in contrast, address large collections of historical provenance data, supporting trend analysis, optimization of future runs, and cross-workflow comparisons. Both modes are essential for a robust provenance management system, as they balance immediate insight with long-term understanding.

A further challenge arises because some workflow managers dynamically expand abstract workflow specifications into many concrete tasks at runtime. Provenance systems must therefore support queries at both levels of abstraction, for example, resolving all concrete executions of a given abstract task, to support meaningful workflow-aware analysis.

\section{HyProv - A Hybrid Workflow Provenance Store}
\label{sec:system}
\subsection{System Overview}

\begin{figure*}[h!]
    \centering
    \includegraphics[width=\textwidth]{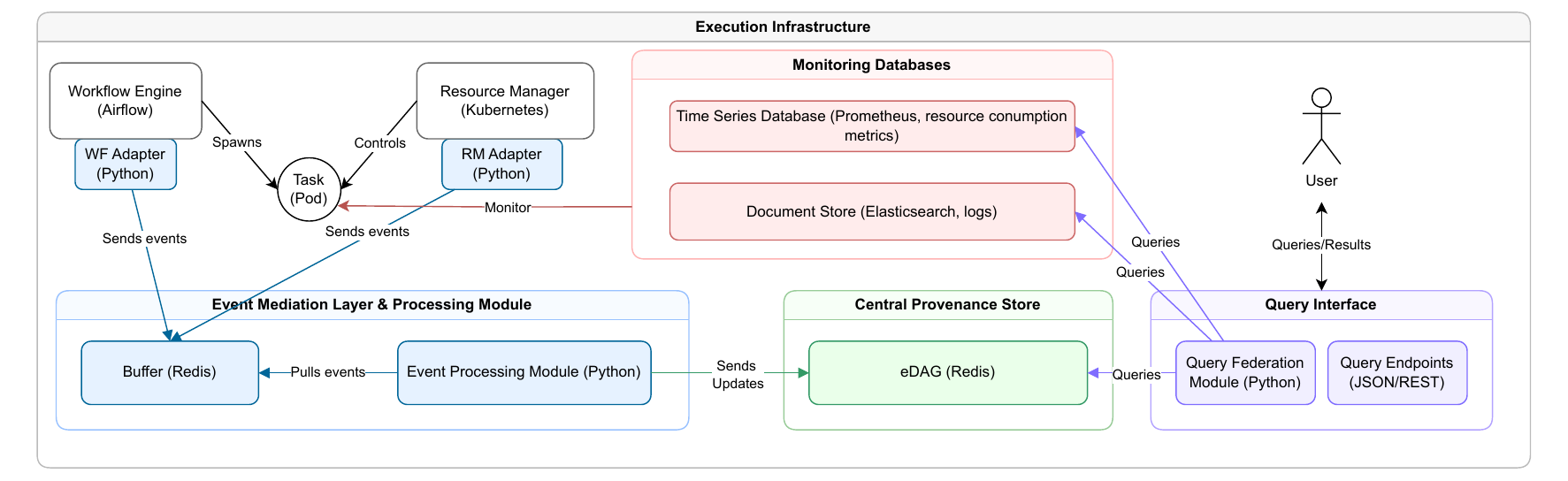}
    \caption{\centering HyProv system architecture, with implementation software noted in parentheses. The core HyProv components—the Event Mediation Layer \& Processing Module (blue), the Central Provenance Store (green), and the Query Interface (purple)—form the core of the system. They integrate with existing infrastructure components, shown in black and red, such as the workflow engine (Airflow), resource manager (Kubernetes), and monitoring databases (Prometheus, Elasticsearch).}
    \label{fig:system_overview}
\end{figure*}

HyProv consolidates selected, distributed provenance information into a single connected model while preserving access to raw data in underlying systems.
For this, HyProv combines a centralized enriched DAG (eDAG) with federated data integration to capture and query workflow provenance across distributed systems (Figure~\ref{fig:system_overview}). 
It builds a central graph that links workflow tasks with their execution infrastructure, while simultaneously federating queries to underlying monitoring and logging systems when needed. 
This hybrid design allows users to query workflow provenance as if it were centralized, without losing the scalability and online benefits of federated data sources.
The architecture is organized around four conceptual components:

\begin{itemize}
    \item \textbf{eDAG}: the central graph-structured model that represents workflow tasks, their dependencies, and execution context.
    \item \textbf{Event Mediation Layer}: connectors to heterogeneous provenance sources that translate raw signals into structured events; this layer also includes an event buffer that decouples producers from consumers and preserves source ordering.
    \item \textbf{Event Processing Module}: a coordinating component that consumes buffered events, applies idempotent updates, and continuously enriches the eDAG.
    \item \textbf{Query Interface}: an external access layer that exposes provenance data through a single set of semantics, supporting both  ``local" queries on the central model and federated queries across distributed stores.
\end{itemize}

% In essence, 
The following sections explain each component in detail. 

\subsection{eDAG model}
The eDAG is derived from the workflow’s execution DAG, where each node represents a concrete task instance and directed edges indicate dependencies between tasks. Nodes are enriched with attributes that link the task to its execution environment—such as the specific execution unit (container or process ID), the cluster node where it ran, start and end timestamps, and current status (e.g., queued, running, succeeded).

For workflows managed by systems with a static DAG (e.g., Airflow or Pegasus), the eDAG can be initialized in full before execution begins, based on the known task graph. In contrast, for workflows with dynamic engines (e.g., Nextflow or Snakemake) where the DAG unfolds at runtime, the eDAG is built incrementally: new task nodes are added as tasks become ready, and they are connected to their predecessors as dependencies resolve. In both cases, the eDAG provides a consistent, up-to-date representation of the workflow execution.

During execution, the eDAG is progressively enriched with events. As tasks transition through states, the event processor updates the corresponding node’s attributes (for example, recording when a task starts or finishes and linking the task to a specific pod and node in the cluster). This continuous enrichment yields a live, integrated view of the workflow and its platform. By capturing links between logical tasks and physical execution units, the eDAG re-integrates context that would otherwise be scattered across workflow logs and infrastructure monitors. The result is a single graph that can be queried for both high-level workflow provenance and low-level execution details.

\subsection{Event Mediation Layer}

This layer connects HyProv to heterogeneous provenance sources and turns their raw signals into loosely \emph{structured events}.

Source–specific \emph{adapters} attach to workflow and infrastructure components (e.g., WFMS scheduler logs, RM/Kubernetes event streams) and perform a lightweight, per–record normalization, harmonize timestamps, and apply minimal de–duplication.
After this normalization step, adapters \emph{emit} the structured events into a \emph{event buffer}.
The buffer decouples event production from consumption, absorbing short–term bursts and regulating flow between components. 
It ensures that the Processing Pipeline operates on uniform events rather than heterogeneous raw logs, while preserving consistent semantics across sources.

\subsection{Event Processing Module}
\label{sec:event-processing-module}

The \emph{Event Processing Module} is the bridge between incoming event streams and the provenance graph. Whereas adapters expose events from heterogeneous sources, the module decides which of these are relevant for provenance and translates them into structured updates. This filtering step is essential, as infrastructure logs typically contain a large volume of low-level noise that is unrelated to workflow execution. 

Relevant events are then categorized into a small set of canonical types (e.g., \texttt{task\_running}, \texttt{task\_succeeded}, \texttt{task\_failed}). For each categorized event, the module extracts the identifiers needed to connect it to the workflow context, such as task names, container instances, node allocations, or abstract tasks. 

The Event Processing Module thus acts as a convergence point: it filters for provenance relevance, normalizes heterogeneous events into a uniform event vocabulary, and applies them to enrich the eDAG. As workflows progress, the graph is continuously updated with task and resource states, providing a coherent and up-to-date view of workflow execution. The raw event streams may not remain available in their original sources. HyProv, on the other hand retains the relevant structured provenance info. By maintaining these associations it enables the connection of data points that would otherwise remain unlinked due to the dynamic and distributed nature of the environment.

\subsection{Query Processing Layer}
\label{sec:query-processing}

The \emph{Query Processing Layer} provides users with a unified way of accessing provenance information. It hides the complexity of multiple data sources behind a single interface and presents queries against the provenance graph as if all relevant information were available in one place. In this way, clients interact with HyProv through a coherent query interface, without needing to know which parts of a query can be answered locally and which require data from external systems.

We distinguish between two types of queries. \emph{Local queries} operate directly on the evolving provenance graph (eDAG). They answer questions about workflow execution that can be derived from the graph alone, such as the state of tasks, their dependencies, or lineage relationships. These queries remain self-contained and do not require access to external databases.

\emph{Federated queries}, by contrast, combine provenance information from the eDAG with additional data from distributed monitoring or infrastructure systems. For example, a query may identify all tasks of a workflow in the eDAG and then retrieve their resource usage metrics from a time-series store. In this case, the eDAG provides the structural and contextual information, while the federated sources contribute performance or system-level details. The Query Processing Layer coordinates these two sides, aligning identifiers and timescales so that the user receives a single, consolidated answer.

In this way, the Query Processing Layer acts as both an access point and an integration point: it makes local provenance immediately accessible while also enabling cross-cutting questions that span multiple systems, all through the same interface.

\section{Implementation}
\label{sec:prototype}

We have developed a prototype of HyProv to validate its design and evaluate its performance. The prototype integrates popular open-source tools for workflow execution, container orchestration, monitoring, and logging, tied together by our provenance capture and query layer. Specifically, we use Apache Airflow as the WMS, Kubernetes as the cluster resource manager, Prometheus for metric monitoring, Elasticsearch for log storage, and a Python-based in-memory graph (backed by Redis) to maintain the eDAG and event queue. The following describes the prototype’s components and how they interact.

\subsection{eDAG in Redis}

In the prototype, the eDAG is implemented directly in Redis, without a separate graph library. Redis maintains two kinds of structures: (i) a set of relationships that capture dependencies between tasks and map abstract tasks to their concrete instances, and (ii) a set of task attributes that store execution-related details such as status, pod name, node assignment, and timestamps.

Table~\ref{tab:edag_node_cpu_intensive_task_1} shows an example of task attributes as stored in Redis for a concrete task instance. In this example, the task node \texttt{cpu\_intensive\_task\_1} records its abstract task identifier, last status update time, execution status, pod name, and the node where it was executed.

\begin{table}[h!]
\begin{center}
\caption{Attributes of an eDAG Node}
\label{tab:edag_node_cpu_intensive_task_1}
\begin{tabular}{|l|l|}
\hline
\textbf{Attribute}           & \textbf{Exemplary value} \\ \hline
Task Name                    & cpu\_intensive\_task\_1 \\ \hline
Abstract Task Name           & cpu\_intensive\_task \\ \hline
Last Status Update           & 2024-12-11 17:08:54.964 UTC \\ \hline
Current Status               & task\_succeeded \\ \hline
Pod Name                     & cpu-intensive-task-1-dff41d38f92a4 \\ \hline
Node Name                    & hu-worker-c25 \\
\hline
\end{tabular}
\end{center}
\end{table}

\subsection{Workflow Management with Airflow}

In our prototype deployment, Apache Airflow orchestrates the workflow’s execution. Airflow is configured to run on Kubernetes, where each task of the workflow is executed in a separate pod.
Airflow’s scheduler generates logs and events as it queues and launches tasks. These include task state transitions (e.g., when a task is queued, started, retried, succeeded, or failed) along with timestamps and task identifiers.

\subsection{Kubernetes as Resource Manager}

We use Kubernetes as the cluster Resource Manager to run the workflow tasks in containers. Kubernetes is responsible for scheduling each containerized task onto the available nodes, handling task isolation and retries, and managing resource allocation according to the cluster’s capacity. As tasks are scheduled and executed, Kubernetes produces a stream of events that we tap into for provenance purposes.

\subsection{Adapters and Event Integration}

Adapters are deployed as sidecar containers alongside the event processor and Redis. This ensures they run close to the data sources they monitor, while remaining loosely coupled to the core provenance logic.

Each adapter is tailored to its source system. The Airflow adapter watches the scheduler logs to detect task state changes and emits events such as \texttt{task\_queued}, \texttt{task\_running}, or \texttt{task\_succeeded}. The Kubernetes adapter subscribes to the event stream of the cluster, reporting provenance-relevant events such as pod assignments, scheduling decisions, or failure notifications.

Events are placed into the Redis queue, from which the central event processor consumes them. 

\subsection{Elasticsearch for Logs}

Elasticsearch stores unstructured and semi-structured logs produced during workflow execution, such as application output, container logs, and system alerts. In the prototype, it provides full-text indexing and fast search capabilities, allowing provenance queries that require keyword filtering or log-based diagnostics.

\subsection{Prometheus for Metrics}

Prometheus captures time-series metrics from the Kubernetes cluster, including CPU, memory, and I/O usage of pods and nodes. These high-frequency measurements are essential for answering federated provenance queries about resource consumption. 

\subsubsection{API Overview}
\label{sec:api}

Each API endpoint is associated with a query template specific to the underlying databases being queried. The templates include placeholders for infrastructure entities. When an endpoint is called, the relevant information is fetched from the eDAG, and the corresponding infrastructure entities are extracted. Afterwards, they are used to fill the placeholders in the query template. Results from the endpoint call are returned as JSON responses. 

The following API endpoints provide access to various task and workflow details, with query parameters allowing for fine-grained filtering based on execution time, task status, and other relevant criteria.

% \textbf{Endpoints for local queries}:
\begin{table}[h!]
\centering
\caption{HyProv API Endpoints and Parameters}
\label{tab:api_endpoints}
\begin{tabular}{|l|l|}
\hline
\textbf{Local Query Endpoints} & \textbf{Description} \\ \hline
\texttt{/get/tasks/\{task\_id\}} & Retrieve task details \\ \hline
\texttt{/get/workflow/nodes} & List all workflow nodes \\ \hline
\texttt{/get/workflow/abstract\_tasks} & Retrieve abstract tasks \\ \hline
\texttt{/get/workflow/tasks} & Retrieve workflow tasks \\ \hline
\texttt{/get/node/tasks/} & List tasks for a node \\ \hline
\end{tabular}

\vspace{0.5em}

\begin{tabular}{|l|l|}
\hline
\textbf{Federated Query Endpoints} & \textbf{Description} \\ \hline
\texttt{/get/tasks/\{task\_id\}/CPU} & CPU usage data \\ \hline
\texttt{/get/tasks/\{task\_id\}/RAM} & RAM usage data \\ \hline
\texttt{/get/tasks/\{task\_id\}/logs} & Task logs \\ \hline
\end{tabular}

\vspace{0.5em}

\begin{tabular}{|l|l|}
\hline
\textbf{Query Parameters} & \textbf{Description} \\ \hline
\texttt{start}, \texttt{end} & Time interval for query \\ \hline
\texttt{task\_status} & Filter by task status \\ \hline
\texttt{abstract\_id} & Abstract task identifier \\ \hline
\texttt{parent\_of}, \texttt{child\_of} & Parent or child task filter \\ \hline
\texttt{node\_id} & Identifier of compute node \\ \hline
\texttt{last\_status\_update} & Last status change timestamp \\ \hline
\texttt{full\_text\_query} & Search string for logs \\ \hline
\end{tabular}
\end{table}

\section{Evaluation}
\label{sec:experiments}
This section describes the experimental setup, including the infrastructure configuration, the workflows used, the queries we will be executing through HyProv. The results of our experiments assess HyProv's workflow-awareness, scalability, ability to work in an online manner, and the additional load it puts on the cluster resources.

We do not include a direct comparison with existing provenance systems for two reasons. 
First, several of our representative queries are not directly expressible in many prior frameworks, which typically separate workflow-level and infrastructure-level provenance. 
Second, comparable systems such as Komadu and Pegasus are primarily designed for HPC and grid environments, while others like Provlake target cloud-native deployments but follow different architectural assumptions. 
While such an evaluation would be valuable, we consider it out of scope for the present work.
% and plan to address it in future studies.

\subsection{Experimental Setup}

The experiments are conducted on a Kubernetes cluster consisting of four homogeneous nodes. System services, including Airflow, Prometheus, Elasticsearch, and the HyProv components, run on dedicated master nodes. Each node is equipped with an Intel Xeon Silver 4314 CPU (16 cores, 32 threads, base frequency 2.40\,GHz, turbo up to 3.40\,GHz). In total, each node provides 32 hardware threads. The nodes have 256\,GB of DDR4 memory, configured as eight 32\,GB ECC DIMMs (3200\,MT/s, operating at 2666\,MT/s). For storage, each node provides local disks alongside a Ceph cluster; however, the synthetic workflows in our experiments do not use the distributed storage backend. 

The cluster runs Kubernetes v1.27.7 as the resource manager, with container orchestration based on containerd. Workflow execution is managed by Apache Airflow v3.0.1 using the CeleryExecutor and the KubernetesPodOperator for tasks. Monitoring data is collected using Prometheus (deployed via Prometheus Operator v0.82.2) with a 10-second scraping interval, while workflow and system logs are stored in Elasticsearch v9.1.0. All nodes run Ubuntu 22.04.4 LTS as the operating system. 

The synthetic workflow is designed to assess the system’s scalability and query processing under controlled conditions. It consists of CPU-intensive, memory-intensive, and combined tasks, each defined with fixed resource profiles. This workflow is executed in three configurations (small: 10 tasks; medium: 100 tasks; large: 1000 tasks) to simulate varying task loads. While large-scale HPC workloads may involve tens or even hundreds of thousands of tasks, many typical scientific workflows consists of hundreds to a few thousand tasks, which our experimental configurations are designed to represent.

\subsection{Workflow-Aware Querying}
We showcase three queries that demonstrate the style of queries HyProv is capable of processing. These queries are representative of typical real-life provenance questions that require combining workflow-level abstractions with runtime and infrastructure information. To ensure comparability, we use synthetic workflows of multiple sizes in our experiments. We do not report performance numbers in this subsection (those results are presented later from Subsection~\ref{sec:scalability} onwards); rather, we use this subsection to motivate and explain the nature and mechanics of the query processing.

Each query demonstrates how HyProv leverages the eDAG structure and, when necessary, external sources such as Prometheus or Elasticsearch to provide answers that would not be possible using individual systems alone. As described in Section~\ref{sec:api}, these queries are executed via the HyProv API, with local queries relying solely on the eDAG, and federated queries involving integration with monitoring systems.

\subsubsection{What node were the failing tasks on?}
%\subsubsection{Identifying problematic nodes}

When executing a large-scale workflow on a cluster, some of the individual task executions may fail. These failures are often associated with problems on the node where they were scheduled. Identifying whether such failures are concentrated on a single node or distributed across the cluster is a common task for workflow managers, as it supports debugging and diagnosis. This information helps determine whether mitigation should focus on isolating a faulty node or on more general rescheduling strategies.
Our objective is therefore to identify the physical nodes on which tasks have failed. We formulate the query to the API endpoint:  
\texttt{/get/workflow/tasks?task\_status=failed}

The endpoint returns all failed concrete tasks, along with their associated metadata, including node assignments and status. The response is then aggregated on the client side into a set of nodes that experienced task failures. This is a local query, since all required data (task status and node identifiers) is already present in the eDAG. No access to external monitoring systems is required.

\subsubsection{What is the CPU and memory usage of this abstract task?}

This query retrieves the resource usage of all task instances corresponding to a specific abstract task. In practice, workflow managers often need to determine whether a given task consistently consumes more or fewer resources than expected. For instance, if a task regularly underutilizes its allocations, requests can be lowered to free cluster capacity, while tasks that exceed expectations may require higher resource requests or scheduling on dedicated nodes. Such queries are therefore essential for informed resource tuning and cluster optimization.

The user issues the following HyProv API calls, providing the abstract task identifier as a query parameter:

\begin{itemize}
    \item \texttt{/get/tasks/\allowbreak CPU?abstract\_task=\allowbreak\{abstract\_task\_id\}} — to retrieve CPU usage;
    \item \texttt{/get/tasks/\allowbreak RAM?abstract\_task=\allowbreak\{abstract\_task\_id\}} — to retrieve memory usage.
\end{itemize}

Upon receiving the query, HyProv first filters the eDAG to identify all concrete task instances associated with the given abstract task. It then resolves each instance to its corresponding execution object (i.e., Kubernetes pod) and extracts the relevant runtime intervals. Finally, having gathered this vital information, it queries Prometheus to retrieve CPU and memory usage data for those pods.

This is a \textbf{federated query}, as it combines logical workflow-level information (abstract-to-concrete resolution) with infrastructure-level metrics stored in an external monitoring system .

\begin{figure}[h!]
  \centering
  % TODO: Replace with actual figure
  \includegraphics[width=0.5\textwidth]{
  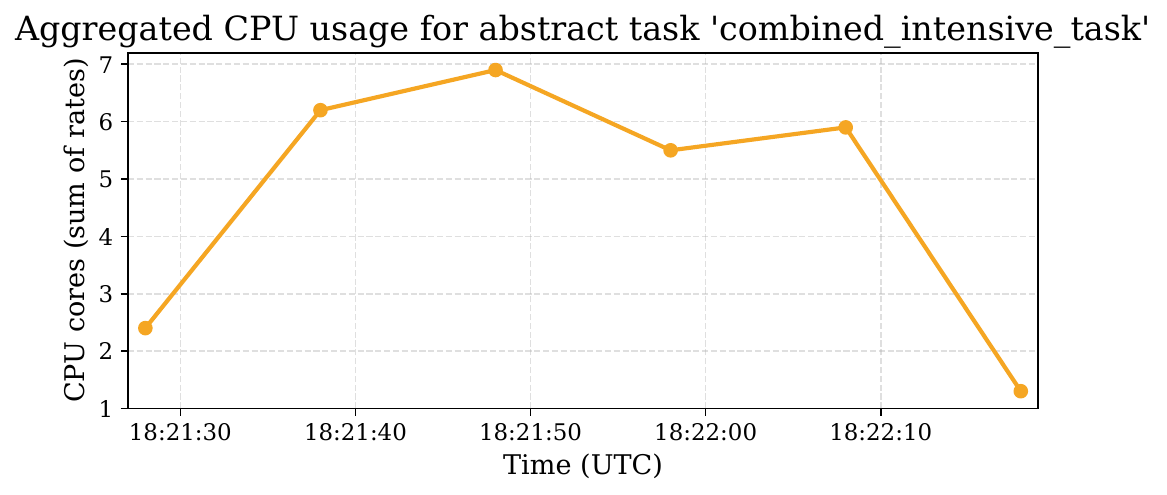}
  \caption{CPU usage for all instances of a specific abstract task.}
  \label{fig:cpu_usage}
\end{figure}

Figure~\ref{fig:cpu_usage} shows an example visualization of the data retrieved by running Query 2 CPU usage across multiple executions of a single abstract task in the medium-size synthetic workflow.

\subsubsection{What tasks that depend on this particular task emitted a warning in their logs?}

This query investigates whether any tasks that are downstream of a given task, i.e., tasks that directly depend on its output, have emitted warning messages during execution. In real workflows, such warnings can indicate partial data issues or degraded performance that may propagate along dependency chains. Detecting them is therefore essential for assessing whether localized problems affect subsequent tasks and for deciding if workflow outputs remain reliable.

To make this query evaluable in a controlled setting, we slightly modify the synthetic workflow so that the ``combined\_intensive'' tasks emit the word \texttt{warning} in their logs. This allows us to simulate warning conditions without relying on actual runtime failures.

To express this query using the HyProv API, the user issues a single call to the task log endpoint, filtered using query parameters:

\begin{itemize}
    \item \texttt{/get/tasks/\allowbreak logs?child\_of=\allowbreak\{task\_id\}\allowbreak\&full\_text\_query=\allowbreak warning}
\end{itemize}

Here, \texttt{child\_of} identifies the dependent tasks via the eDAG, and \texttt{full\_text\_query} restricts the result to those that emitted warning messages in their logs.

HyProv processes this query by resolving dependencies through the eDAG information stored in Redis, where each task edge captures dependency links. For the relevant tasks, it gets relevant data (pod, time information, status). With this information, and the information provided in the API endpoint, it builds and issues a single query to Elasticsearch.

This is again a \textbf{federated query}, combining workflow-level dependency resolution with external log inspection. It highlights HyProv’s ability to unify structural provenance with runtime observability in a single, declarative query.

\subsection{Scalability Analysis}
\label{sec:scalability}

This section evaluates HyProv’s ability to scale with increasing workflow sizes by measuring query response times across small, medium, and large configurations of the synthetic workflow. We benchmark all three representative queries presented earlier, each executed three times per configuration. To ensure consistency, the system cache is cleared between runs, and query response times are measured from the moment the request is issued until the response is fully retrieved.

\subsubsection{What node were the failing tasks on?}
Table~\ref{tab:online_query_times} presents the results for the first query, which retrieves the set of failed tasks and the nodes they ran on. For this query, the reported time reflects the duration required to collect the full list of failed tasks, including their associated node identifiers, as returned by the eDAG. The final step of aggregating or formatting the set of nodes is performed client-side and is not included in the measured query time.

As explained above, this is a local query, executed entirely on top of the in-memory Redis store. Expectedly, query times scale much more favorably than the number of tasks: from 10 to 1000 tasks, the average response time increases from ~6 ms to ~129 ms. This non-linear growth reflects the efficiency of Redis-based querying and the absence of external system calls.

\begin{table}[h!]
\centering
\caption{Query Response Times for Query 1 (Local Query: Failed Tasks with Node Info)}
\label{tab:online_query_times}
\begin{tabular}{|l|c|c|c|c|}
\hline
\textbf{Configuration} & \textbf{\# of Tasks} & \textbf{Min (ms)} & \textbf{Avg (ms)} & \textbf{Max (ms)} \\ \hline
Small & 10 & 5.704 & 5.768 & 5.837 \\ \hline
Medium & 100 & 13.419 & 14.507 & 16.539 \\ \hline
Large & 1000 & 119.538 & 124.338 & 129.149 \\ \hline
\end{tabular}
\end{table}

\subsubsection{What is the CPU and memory usage of this abstract task?}

The second scalability experiment measures the time required to execute the federated query that retrieves CPU usage for all instances of a given abstract task. This query was introduced earlier and relies on Prometheus as the backend for time-series metrics.

HyProv internally resolves the abstract task to all concrete task instances and their runtime intervals, then issues a single time-bounded PromQL query to Prometheus to compute CPU usage over time.

Table~\ref{tab:cpu_query_times} presents the response times for this query. Unlike the local query in the previous experiment, the response time here reflects network latency and the cost of evaluating the PromQL expression over the selected time range. The scaling remains sub-linear thanks to HyProv’s in-memory  resolution of execution metadata and the use of a single aggregated query per metric type.

\begin{table}[h!]
\centering
\caption{Query Response Times for Query 2 (Federated Query: CPU Usage of Abstract Task)}
\label{tab:cpu_query_times}
\begin{tabular}{|l|c|c|c|c|}
\hline
\textbf{Configuration} & \textbf{\# of Tasks} & \textbf{Min (ms)} & \textbf{Avg (ms)} & \textbf{Max (ms)} \\ \hline
Small & 10 & 11.1 & 11.7 & 12.2 \\ \hline
Medium & 100 & 17.0 & 19.9 & 21.9 \\ \hline
Large & 1000 & 104.0 & 115.8 & 126.3 \\ \hline
\end{tabular}
\end{table}

\subsubsection{What tasks that depend on this particular task emitted a warning in their logs?}

The third scalability experiment measures the time required to execute the federated query that identifies which downstream tasks of a given task emitted warning messages during execution. As described earlier, HyProv first resolves all direct child tasks via the eDAG using its knowledge of the DAG, and then queries Elasticsearch for log entries containing the string \texttt{warning} for the execution objects resolved in the previous step. To create controlled test conditions, we modify the synthetic workflow so that some tasks emit the word \texttt{warning} in their logs , simulating fault-like behavior without relying on actual runtime errors.

The measured times include both the dependency resolution in Redis and the subsequent log search in Elasticsearch. Table~\ref{tab:warning_query_times} presents the results. While response times are naturally higher than for purely local queries, they remain in the range of hundreds of milliseconds even for the large configuration. 

\begin{table}[h!]
\centering
\caption{Query Response Times for Query 3 (Federated Query: Downstream Tasks Emitting Warnings)}
\label{tab:warning_query_times}
\begin{tabular}{|l|c|c|c|c|}
\hline
\textbf{Configuration} & \textbf{\# of Tasks} & \textbf{Min (ms)} & \textbf{Avg (ms)} & \textbf{Max (ms)} \\ \hline
Small & 10 & 57.3 & 65.3 & 72.6 \\ \hline
Medium & 100 & 65.6 & 68.0 & 71.2 \\ \hline
Large & 1000 & 122.2 & 125.3 & 129.5 \\ \hline
\end{tabular}
\end{table}

\subsection{Online Querying}
\label{sec:online_querying}

In this subsection, we evaluate HyProv’s \emph{online} capabilities. Because Prometheus and Elasticsearch are widely used and highly optimized backends, we treat their internal query engines as black boxes and focus instead on HyProv’s contribution to end-to-end freshness. Specifically, we measure the \emph{ingestion latency}, defined as the time from when an external event occurs to when the corresponding provenance state becomes available in the eDAG.

We conducted five independent runs for each workflow size (small, medium, and large) to measure HyProv’s ingestion latency. For each event, we record the time of occurrence at the source (Airflow or Kubernetes) and the time the corresponding update becomes available in the eDAG. We then report the distribution of these latencies using the median (p50), 95th percentile (p95), and the maximum time across all observed events. These values provide a representative view of typical ingestion delay as well as the system’s behavior under load in the latency tail.

Table~\ref{tab:ingestion_latency} summarizes the results. Latencies scale modestly with workflow size, growing sub-linearly with the number of tasks, and even at the largest configuration, the tail latencies (p95) remain comfortably below one quarter of a second.

\begin{table}[h!]
\centering
\caption{Ingestion Latency from Event Observation to eDAG Availability}
\label{tab:ingestion_latency}
\begin{tabular}{|l|c|c|c|c|c|}
\hline
\textbf{Configuration} & \textbf{\# of Events} & \textbf{p50 (ms)} & \textbf{p95 (ms)}  & \textbf{Max (ms)} \\ \hline
Small  & 250 & 1.992 & 3.693 & 5.074 \\ \hline
Medium & 2265 & 3.471 & 24.963 & 38.161 \\ \hline
Large  & 24185 & 23.087 & 130.683 & 174.045 \\ \hline
\end{tabular}
\end{table}

\subsection{Overhead Measurements}
To quantify HyProv’s resource overhead, we executed the large synthetic workflow—the most demanding configuration—five times and measured CPU and memory consumption across all core components (adapters, Redis, eDAG, and processing). Table~\ref{tab:overhead_summary} reports the minimum, average, and maximum utilization observed across runs. CPU demand remained very low, averaging only 0.06 cores and never exceeding 0.23 cores, while memory consumption was stable at around 1.46\,GiB, with only minor fluctuations between 1.43\,GiB and 1.48\,GiB. These findings demonstrate that HyProv imposes only limited overhead even under the most resource-intensive workload.

\begin{table}[h!]
\centering
\caption{HyProv Overhead on Large Synthetic Workflow (aggregated across five runs)}
\label{tab:overhead_summary}
\begin{tabular}{|l|c|c|c|}
\hline
\textbf{Metric} & \textbf{Min} & \textbf{Avg} & \textbf{Max} \\ \hline
CPU (cores)   & 0.005 & 0.064 & 0.229 \\ \hline
Memory (GiB)  & 1.426 & 1.465 & 1.481 \\ \hline
\end{tabular}
\end{table}

\section{Conclusion}
\label{sec:conclusion}
This paper introduced HyProv, a hybrid provenance management system that makes workflow executions first-class citizens by enabling workflow-aware queries across both workflow and infrastructure layers. Unlike centralized approaches, HyProv combines online processing with a federated design that stores only compact eDAG execution data while reusing existing provenance and monitoring systems. This architecture ensures scalability and responsiveness without duplicating large volumes of data. Our evaluation confirms that HyProv remains lightweight in practice: even under the most demanding workflow configuration, CPU usage of the processing components averaged only 0.06 cores and never exceeded 0.23 cores, while memory consumption stayed stable around 1.46\,GiB. Moreover, the system scales sublinearly with workflow size, maintaining modest overhead even as the number of tasks and events increases exponentially. Together, these results demonstrate that HyProv provides workflow-aware, scalable, and online provenance management with limited resource costs.

Future work may explore how the collected metadata can support the optimization of external provenance and monitoring databases, for example, through improved indexing or query planning—particularly for queries that span multiple federated systems. It would also be valuable to assess HyProv’s extensibility by integrating additional provenance sources beyond Kubernetes (e.g., storage or filesystem services) and by applying it to further workflow management systems and infrastructure backends. Lastly, a systematic comparison with other provenance approaches—both centralized and hybrid—would be a valuable direction for future work, helping to clarify the trade-offs in design, deployment complexity, and query expressiveness.

\section*{Acknowledgments}
We acknowledge funding from the Deutsche Forschungsgemeinschaft through the SFB 1404 FONDA (Project-ID 414984028).

\vspace{12pt}


\begin{thebibliography}{00}
\bibitem{souzaLightweightDataIntegration2023a} Souza, R., Skluzacek, T., Wilkinson, S., Ziatdinov, M. \& Da Silva, R. Towards Lightweight Data Integration Using Multi-Workflow Provenance and Data Observability. {\em 2023 IEEE 19th International Conference On E-Science (e-Science)}. 2023.

\bibitem{ditommasoNextflowEnablesReproducible2017} Di Tommaso, P., Chatzou, M., Floden, E., Barja, P., Palumbo, E. \& Notredame, C. Nextflow Enables Reproducible Computational Workflows. {\em Nature Biotechnology}. 2017.

\bibitem{papadimitriouEndendOnlinePerformance2021} Papadimitriou, G., Wang, C., Vahi, K., Da Silva, R., Mandal, A., Liu, Z., Mayani, R., Rynge, M., Kiran, M., Lynch, V., Kettimuthu, R., Deelman, E., Vetter, J. \& Foster, I. End-to-End Online Performance Data Capture and Analysis for Scientific Workflows. {\em Future Generation Computer Systems}. 2021.

\bibitem{souzaEfficientRuntimeCapture2019a} Souza, R., Azevedo, L., Thiago, R., Soares, E., Nery, M., Netto, M., Vital, E., Cerqueira, R., Valduriez, P. \& Mattoso, M. Efficient Runtime Capture of Multiworkflow Data Using Provenance. {\em 2019 15th International Conference On E-Science (e-Science)}. 2019.

\bibitem{vahiGeneralApproachRealTime2012} Vahi, K., Harvey, I., Samak, T., Gunter, D., Evans, K., Rogers, D., Taylor, I., Goode, M., Silva, F., Al-Shakarchi, E., Mehta, G., Jones, A. \& Deelman, E. A General Approach to Real-Time Workflow Monitoring. {\em 2012 SC Companion: High Performance Computing, Networking Storage And Analysis}. 2012.

\bibitem{suriarachchiKomaduCaptureVisualization2015} Suriarachchi, I., Zhou, Q. \& Plale, B. Komadu: A Capture and Visualization System for Scientific Data Provenance. {\em Journal Of Open Research Software}. 2015.

\bibitem{gehaniEfficientQueryingDistributed2010} Gehani, A., Kim, M. \& Malik, T. Efficient Querying of Distributed Provenance Stores. {\em Proceedings Of The 19th ACM International Symposium On High Performance Distributed Computing}. 2010.

\bibitem{ellqvistUsingMediationAchieve2009} Ellqvist, T., Koop, D., Freire, J., Silva, C. \& Strömbäck, L. Using Mediation to Achieve Provenance Interoperability. {\em 2009 Congress On Services - I}. 2009.

\bibitem{perezSystematicReviewProvenance2018a} Pérez, B., Rubio, J. \& Sáenz-Adán, C. A Systematic Review of Provenance Systems. {\em Knowledge And Information Systems}. 2018.

\bibitem{yatesReproduciblePortableEfficient2021} Yates, J., Lamnidis, T., Borry, M., Valtueña, A., Fagernäs, Z., Clayton, S., Garcia, M., Neukamm, J. \& Peltzer, A. Reproducible, Portable, and Efficient Ancient Genome Reconstruction with Nf-Core/Eager. {\em PeerJ}. 2021.

\bibitem{lehmannFORCENextflowScalable} Lehmann, F., Frantz, D., Becker, S., Leser, U. \& Hostert, P. FORCE on Nextflow: Scalable Analysis of Earth Observation Data on Commodity Clusters. {\em Proceedings Of The CIKM 2021 Workshops}. 2021.

\bibitem{schaarschmidtWorkflowEngineeringMaterials2022} Schaarschmidt, J., Yuan, J., Strunk, T., Kondov, I., Huber, S., Pizzi, G., Kahle, L., Bölle, F., Castelli, I., Vegge, T., Hanke, F., Hickel, T., Neugebauer, J., Rêgo, C. \& Wenzel, W. Workflow Engineering in Materials Design within the BATTERY 2030+ Project. {\em Advanced Energy Materials}. 2022.

\bibitem{liewScientificWorkflowsMoving2016} Liew, C., Atkinson, M., Galea, M., Ang, T., Martin, P. \& Hemert, J. Scientific Workflows: Moving Across Paradigms. {\em ACM Computing Surveys}. 2016.

\bibitem{khanSharingInteroperableWorkflow2019} Khan, F., Soiland-Reyes, S., Sinnott, R., Lonie, A., Goble, C. \& Crusoe, M. Sharing Interoperable Workflow Provenance: A Review of Best Practices and Their Practical Application in CWLProv. {\em GigaScience}. 2019.

\bibitem{davidsonProvenanceScientificWorkflows} Davidson, S. \& Freire, J. Provenance and Scientific Workflows: Challenges and Opportunities. {\em Proceedings Of The ACM SIGMOD International Conference On Management Of Data (SIGMOD)}. 2008.

\bibitem{herschelSurveyProvenanceWhat2017b} Herschel, M., Diestelkämper, R. \& Ben Lahmar, H. A Survey on Provenance: What for? What Form? What From?. {\em The VLDB Journal}. 2017.

\bibitem{deelmanPegasusWorkflowManagement2015} Deelman, E., Vahi, K., Juve, G., Rynge, M., Callaghan, S., Maechling, P., Mayani, R., Chen, W., Ferreira da Silva, R., Livny, M. \& Wenger, K. Pegasus, a Workflow Management System for Science Automation. {\em Future Generation Computer Systems}. \textbf{46}, 2015.

\bibitem{harrisonWSRFWorkflowTriana2008} Harrison, A., Taylor, I., Wang, I. \& Shields, M. WS-RF Workflow in Triana. {\em International Journal Of High Performance Computing Applications}. 2008.

\bibitem{deelmanWorkflowsEScienceOverview2009} Deelman, E., Gannon, D., Shields, M. \& Taylor, I. Workflows and E-Science: An Overview of Workflow System Features and Capabilities. {\em Future Generation Computer Systems}. 2009.

\bibitem{liuSurveyDataIntensiveScientific2015a} Liu, J., Pacitti, E., Valduriez, P. \& Mattoso, M. A Survey of Data-Intensive Scientific Workflow Management. {\em Journal Of Grid Computing}. 2015.

\bibitem{buxParallelizationScientificWorkflow2013} Bux, M. \& Leser, U. Parallelization in Scientific Workflow Management Systems. {\em ArXiv:1303.7195 [cs]}. 2013.

\end{thebibliography}
\end{document}